\documentclass[aps,prd,prabib,showpacs,nofootinbib,10pt]{revtex4}
\usepackage{graphicx} \usepackage{amsmath} \usepackage{amssymb}
\usepackage{amsfonts} \usepackage{bm}
\usepackage{array}
\usepackage{siunitx}
\usepackage[singlelinecheck=false]{caption}
\usepackage{ytableau}
\usepackage{multirow}
\usepackage{mathtools}

\DeclarePairedDelimiter\floor{\lfloor}{\rfloor}
\usepackage{tikz}
\usetikzlibrary{calc}

\begin{document}

\newcommand{\be}{\begin{equation}} \newcommand{\ee}{\end{equation}}
\newcommand{\bea}{\begin{eqnarray}}\newcommand{\eea}{\end{eqnarray}}

\title{Quantum  search  on Hanoi network}

\author{Pulak Ranjan Giri} \email{pulakgiri@gmail.com}

\affiliation{International Institute of Physics, Universidade Federal do Rio Grande do Norte, Campus Universitario, Lagoa Nova, Natal-RN 59078-970, Brazil} 

\author{Vladimir   Korepin} \email{korepin@gmail.com}

\affiliation{C. N. Yang Institute for Theoretical Physics, State University of New York at  Stony Brook, Stony Brook, NY 11794-3840, USA}

\begin{abstract} 
Hanoi network  has  a one-dimensional periodic lattice  as  its main structure with additional  long-range edges, which allow  having  efficient  quantum walk algorithm  that  can   find a target state on  the network    faster than the exhaustive  classical search.   In this article, we  use     regular quantum walks   and  lackadaisical quantum walks    respectively   to   search for a target state.    From the curve  fitting  of   the   numerical  results    for  Hanoi network  of degree three and four we  find that  their  running  time for the regular quantum walks  followed by amplitude amplification    scales  as  $\mathcal{O}\left(N^{0.79} \sqrt{\log N}\right)$   and     $\mathcal{O}\left(N^{0.65} \sqrt{\log N}\right)$  respectively.   And    for the  search by lackadaisical quantum walks   the   running time scales  as     $\mathcal{O}\left(N^{0.57}\log N\right)$   and   $\mathcal{O}\left(N^{0.50}\log N\right)$  respectively.

\end{abstract}


\date{\today}

\maketitle 



\section{Introduction} \label{in}
Searching an unsorted database for a target state  has an optimal speed  of $\mathcal{O}(\sqrt{N})$ in quantum computations, which is achieved by the famous Grover  search \cite{grover1, grover2,radha1,korepin1,giri,zhang}. In this algorithm elements of   the database of size,   $N$  are    associated with an orthonormal   basis states  of  an  $N$-dimensional  Hilbert space. 
An initial state  is prepared which has typically a small overlap of  $\mathcal{O}(1/\sqrt{N})$   with the target state. Then amplitude amplification technique is exploited to evolve the initial state to the target state in  $\mathcal{O}(\sqrt{N})$  oracle queries, which is quadratically faster than the best known exhaustive classical search of time complexity  $\mathcal{O}(N)$.  The efficiency of quantum computations   over its classical counterpart  was anticipated over three and half decades ago by  Richard Feynman and  Paul Benioff.  Although Grover algorithm does not have exponential speedup  in comparison to the classical algorithm it has become  very much  popular since  it can be used as a subroutine in   numerous  applications.

The transition from combinatorial search problem such as Grover   to    searching  a spatially distributed   database  seemed difficult at the beginning.  As pointed out by Paul Benioff \cite{beni}, it will take time for a quantum robot to move from one vertex to the other eventually losing the  efficiency over classical search. 
Naive implementation    of  Grover algorithm  on a  2-dimensional lattice   requires     $\mathcal{O}(\sqrt{N})$   oracle calls, but  each Grover  iteration   needs  $\mathcal{O}(\sqrt{N})$  time   steps  for the reflection  operation   making  the total running time     $\mathcal{O}(N)$, which is   not    better than    classical  search.  Usually,  physical database is  modeled as  a graph where   vertices \cite{childs,amba1,amba2,meyer,amba4}  are considered  as the elements of the database.    It is possible to  search  this   graph   for a target state  faster than the   classical   search.

For example, one-dimensional periodic  lattice can be searched for a target state in   $\mathcal{O} (N)$  time steps \cite{giri1} with  $\mathcal{O}(1)$ success probability by lackadaisical quantum walk followed by amplitude amplification.  Although,  regular  quantum walk(without laziness) search  in one-dimensional lattice is not efficient.  In two-dimensional periodic lattice  recursive  algorithm  followed by   amplitude amplification  can search for a target state in $\mathcal{O}\left(\sqrt{N}\log^2 N\right)$ time \cite{am,amba1}.  Regular quantum walk followed by amplitude amplification can improve this running time to
$\mathcal{O} \left(\sqrt{N} \log N \right)$  and  Tulsi's  controlled quantum walks \cite{tulsi}, lackadaisical quantum walks \cite{wong1} or other method \cite{amba3} can further improve it to  $\mathcal{O}(\sqrt{N \log N })$.  For $d$-dimensional lattice($d \geq 3$) optimal  speed of  $\mathcal{O}(\sqrt{N})$   \cite{amba2,childs1} 
is achieved by quantum walk  search algorithm.
Other examples of  graphs  where  search by quantum walks   have  been successfully exploited   include   a   complete graph with  directed self-loops   \cite{portugal},  Sierpinski gasket
 \cite{patel},  Hanoi network(HN)  \cite{portugal3,portugal4},  hypercubes,   hexagonal network \cite{portugal1}, triangular lattice \cite{portugal2}   etc.

Hanoi network  is interesting because it has  one-dimensional periodic lattice as its  main structure together with additional long-range edges.  We know that  quantum walks  for  search on  the  one-dimensional periodic lattice is not efficient.   However, the additional long-range  edges of the Hanoi network  make  the   search by quantum walk   efficient  as shown in refs.    \cite{portugal3,portugal4}.   They used  quantum walks followed by amplitude amplification   with   a  modified  Grover coin  which can control the flow of probability flux  on  and outside  of the main structure by  a single parameter.     Running times obtained   by this modified method   for  HN3 and HN4   are     $\mathcal{O}(N^{0.74})$   and  $\mathcal{O}(N^{0.84})$ respectively.    They also exploited Tulsi's  controlled  quantum walks with regular Grover coin which provides   running times  of     $\mathcal{O}(N^{0.62})$   and  $\mathcal{O}(N^{0.65})$ respectively.   Note that, the    lackadaisical quantum walks  followed by amplitude amplification  can search for a target state on a  one-dimensional periodic lattice.  Therefore,  it would be interesting to know how fast  lackadaisical quantum walk can search for a target state on Hanoi  network,  which has  more edges than the one-dimensional periodic lattice.

So, the purpose of this article is to study   spatial search for  a target state  on Hanoi network  with the help of   recently  introduced   lackadaisical(lazy)  quantum walks   \cite{wong1,wong2,wong3},    which  can  improve the  time complexity significantly.   We also want  to study  quantum walk search with  standard  Grover coin. 
This article  is  arranged in the following fashion.  First,  we   discuss  the   Hanoi network   in section \ref{hanoi}.  Then   in section \ref{qw} we  discuss   quantum walk search with standard Grover coin  in this network.   In  section   \ref{lack}   lackadaisical quantum walk  search on Hanoi network is discussed. 
We  conclude in section  \ref{con} with a discussion.

\begin{figure}[h!]
\centering
\begin{tikzpicture}

\def \n {16}
\def \radius {2.0cm}
\draw[thick ]  circle (\radius);

\node[draw, circle, fill = black, scale=0.4, label = right:{\scriptsize$8$}] at ({360/\n * 0}:\radius) {};
\node[draw, circle, fill = black, scale=0.4, label = right:{\scriptsize $7$}] at ({360/\n * 1}:\radius) {};
\node[draw, circle, fill = black, scale=0.4, label = above:{\scriptsize$6$}] at ({360/\n * 2}:\radius) {};
\node[draw, circle, fill = black, scale=0.4, label = above:{\scriptsize$5$}] at ({360/\n * 3}:\radius) {};
\node[draw, circle, fill = black, scale=0.4, label = above:{\scriptsize$4$}] at ({360/\n * 4}:\radius) {};
\node[draw, circle, fill = black, scale=0.4, label = above:{\scriptsize$3$}] at ({360/\n * 5}:\radius) {};
\node[draw, circle, fill = black, scale=0.4, label = left:{\scriptsize$2$}] at ({360/\n * 6}:\radius) {};
\node[draw, circle, fill = black, scale=0.4, label = left:{\scriptsize$1$}] at ({360/\n * 7}:\radius) {};
 \node[draw, circle, fill = black, scale=0.4, label = left:{\scriptsize$16$}] at ({360/\n * 8}:\radius) {};
\node[draw, circle, fill = black, scale=0.4, label = left:{\scriptsize$15$}] at ({360/\n * 9}:\radius) {};
\node[draw, circle, fill = black, scale=0.4, label = below:{\scriptsize$14$}] at ({360/\n * 10}:\radius) {};
\node[draw, circle, fill = black, scale=0.4, label = below:{\scriptsize$13$}] at ({360/\n * 11}:\radius) {};
\node[draw, circle, fill = black, scale=0.4, label = below:{\scriptsize$12$}] at ({360/\n * 12}:\radius) {};
\node[draw, circle, fill = black, scale=0.4, label = below:{\scriptsize$11$}] at ({360/\n * 13}:\radius) {};
\node[draw, circle, fill = black, scale=0.4, label = right:{\scriptsize$10$}] at ({360/\n * 14}:\radius) {};
\node[draw, circle, fill = black, scale=0.4, label = right:{\scriptsize$9$}] at ({360/\n * 15}:\radius) {};

\draw[thick, blue] ({360/\n * 7}:\radius) to [out=-30,in=-70] ({360/\n * 5}:\radius); 
\draw[thick, blue] ({360/\n * 3}:\radius) to [out=-100,in=-170] ({360/\n * 1}:\radius); 
\draw[thick, blue] ({360/\n * 15}:\radius) to [out=170,in=100] ({360/\n * 13}:\radius); 
\draw[thick, blue] ({360/\n * 11}:\radius) to [out=80,in=20] ({360/\n * 9}:\radius); 
\draw[thick, blue] ({360/\n * 6}:\radius) to [out=90,in=90, looseness=1.5] ({360/\n * 2}:\radius); 
\draw[thick, blue] ({360/\n * 14}:\radius) to [out=-90,in=-90, looseness=1.6] ({360/\n * 10}:\radius); 

 \draw[red] ({360/\n * 0}:\radius) to [out= 170,in=-90, loop ] ({360/\n * 0}:\radius);
 \draw[red] ({360/\n * 1}:\radius) to [out= 170,in=-90,  loop] ({360/\n * 1}:\radius);
 \draw[red] ({360/\n * 2}:\radius) to [out= 170,in=-90,  loop] ({360/\n * 2}:\radius);
 \draw[red] ({360/\n * 3}:\radius) to [out= 170,in=-90,  loop ] ({360/\n * 3}:\radius);
 \draw[red] ({360/\n * 4}:\radius) to [out= 170,in=-90,  loop] ({360/\n * 4}:\radius);
 \draw[red] ({360/\n * 5}:\radius) to [out= 170,in=-90,  loop ] ({360/\n * 5}:\radius);
 \draw[red] ({360/\n * 6}:\radius) to [out= 170,in=-90,  loop ] ({360/\n * 6}:\radius);
 \draw[red] ({360/\n * 7}:\radius) to [out= 170,in=-90,  loop ] ({360/\n * 7}:\radius);
 \draw[red] ({360/\n * 8}:\radius) to [out= 170,in=-90,  loop ] ({360/\n * 8}:\radius);
  \draw[red] ({360/\n * 9}:\radius) to [out= 170,in=-90,  loop] ({360/\n * 9}:\radius);
 \draw[red] ({360/\n * 10}:\radius) to [out= 170,in=-90,  loop] ({360/\n * 10}:\radius);
 \draw[red] ({360/\n * 11}:\radius) to [out= 170,in=-90,  loop] ({360/\n * 11}:\radius);
 \draw[red] ({360/\n * 12}:\radius) to [out= 170,in=-90,  loop] ({360/\n * 12}:\radius);
 \draw[red] ({360/\n * 13}:\radius) to [out= 170,in=-90,  loop] ({360/\n * 13}:\radius);
 \draw[red] ({360/\n * 14}:\radius) to [out= 170,in=-90,  loop] ({360/\n * 14}:\radius);
 \draw[red] ({360/\n * 15}:\radius) to [out= 170,in=-90,  loop] ({360/\n * 15}:\radius);
 
 \draw[thick, blue] ({360/\n * 4}:\radius) to [out=-90,in=90, looseness=0.4] ({360/\n * 12}:\radius);

\draw[thick, blue] ({360/\n * 8}:\radius) to [out= 170,in=-90,  loop right ] ({360/\n * 8}:\radius);
\draw[thick, blue] ({360/\n * 0}:\radius) to [out= 170,in=-90,  loop right ] ({360/\n * 0}:\radius);

\node at ({360/\n * 12}:3cm)  {(\textit{a})};
\end{tikzpicture}
\hspace{1cm}
\begin{tikzpicture}
\def \n {16}
\def \radius {2.0cm}
\draw[thick]  circle (\radius);

\node[draw, circle, fill = black, scale=0.4, label = right:{\scriptsize$8$}] at ({360/\n * 0}:\radius) {};
\node[draw, circle, fill = black, scale=0.4, label = right:{\scriptsize $7$}] at ({360/\n * 1}:\radius) {};
\node[draw, circle, fill = black, scale=0.4, label = above:{\scriptsize$6$}] at ({360/\n * 2}:\radius) {};
\node[draw, circle, fill = black, scale=0.4, label = above:{\scriptsize$5$}] at ({360/\n * 3}:\radius) {};
\node[draw, circle, fill = black, scale=0.4, label = above:{\scriptsize$4$}] at ({360/\n * 4}:\radius) {};
\node[draw, circle, fill = black, scale=0.4, label = above:{\scriptsize$3$}] at ({360/\n * 5}:\radius) {};
\node[draw, circle, fill = black, scale=0.4, label = left:{\scriptsize$2$}] at ({360/\n * 6}:\radius) {};
\node[draw, circle, fill = black, scale=0.4, label = left:{\scriptsize$1$}] at ({360/\n * 7}:\radius) {};
 \node[draw, circle, fill = black, scale=0.4, label = left:{\scriptsize$16$}] at ({360/\n * 8}:\radius) {};
\node[draw, circle, fill = black, scale=0.4, label = left:{\scriptsize$15$}] at ({360/\n * 9}:\radius) {};
\node[draw, circle, fill = black, scale=0.4, label = below:{\scriptsize$14$}] at ({360/\n * 10}:\radius) {};
\node[draw, circle, fill = black, scale=0.4, label = below:{\scriptsize$13$}] at ({360/\n * 11}:\radius) {};
\node[draw, circle, fill = black, scale=0.4, label = below:{\scriptsize$12$}] at ({360/\n * 12}:\radius) {};
\node[draw, circle, fill = black, scale=0.4, label = below:{\scriptsize$11$}] at ({360/\n * 13}:\radius) {};
\node[draw, circle, fill = black, scale=0.4, label = right:{\scriptsize$10$}] at ({360/\n * 14}:\radius) {};
\node[draw, circle, fill = black, scale=0.4, label = right:{\scriptsize$9$}] at ({360/\n * 15}:\radius) {};

\draw[thick, blue] ({360/\n * 7}:\radius) to [out=-30,in=-70] ({360/\n * 5}:\radius); 
\draw[thick, blue] ({360/\n * 5}:\radius) to [out=-70,in=-100] ({360/\n * 3}:\radius); 
\draw[thick, blue] ({360/\n * 3}:\radius) to [out=-100,in=-170] ({360/\n * 1}:\radius); 
\draw[thick, blue] ({360/\n * 1}:\radius) to [out=-170,in=170] ({360/\n * 15}:\radius); 
\draw[thick, blue] ({360/\n * 15}:\radius) to [out=170,in=100] ({360/\n * 13}:\radius); 
\draw[thick, blue] ({360/\n * 13}:\radius) to [out=100,in=80] ({360/\n * 11}:\radius); 
\draw[thick, blue] ({360/\n * 11}:\radius) to [out=80,in=20] ({360/\n * 9}:\radius); 
\draw[thick, blue] ({360/\n * 9}:\radius) to [out=20,in=-20] ({360/\n * 7}:\radius); 
\draw[thick, blue] ({360/\n * 6}:\radius) to [out=90,in=90, looseness=1.5] ({360/\n * 2}:\radius); 
\draw[thick, blue] ({360/\n * 2}:\radius) to [out=10,in=-10, looseness=1.6] ({360/\n * 14}:\radius); 
\draw[thick, blue] ({360/\n * 14}:\radius) to [out=-90,in=-90, looseness=1.6] ({360/\n * 10}:\radius); 
\draw[thick, blue] ({360/\n * 10}:\radius) to [out=-170,in=170, looseness=1.6] ({360/\n * 6}:\radius); 
\draw[thick, blue] ({360/\n * 4}:\radius) to [out=-50,in=50, looseness=0.4] ({360/\n * 12}:\radius); 
\draw[thick, blue] ({360/\n * 4}:\radius) to [out=-130,in=140, looseness=0.4] ({360/\n * 12}:\radius); 

 
 \draw[red] ({360/\n * 0}:\radius) to [out= 170,in=-90, loop ] ({360/\n * 0}:\radius);
 \draw[red] ({360/\n * 1}:\radius) to [out= 170,in=-90,  loop] ({360/\n * 1}:\radius);
 \draw[red] ({360/\n * 2}:\radius) to [out= 170,in=-90,  loop] ({360/\n * 2}:\radius);
 \draw[red] ({360/\n * 3}:\radius) to [out= 170,in=-90,  loop ] ({360/\n * 3}:\radius);
 \draw[red] ({360/\n * 4}:\radius) to [out= 170,in=-90,  loop] ({360/\n * 4}:\radius);
 \draw[red] ({360/\n * 5}:\radius) to [out= 170,in=-90,  loop ] ({360/\n * 5}:\radius);
 \draw[red] ({360/\n * 6}:\radius) to [out= 170,in=-90,  loop ] ({360/\n * 6}:\radius);
 \draw[red] ({360/\n * 7}:\radius) to [out= 170,in=-90,  loop ] ({360/\n * 7}:\radius);
 \draw[red] ({360/\n * 8}:\radius) to [out= 170,in=-90,  loop ] ({360/\n * 8}:\radius);
  \draw[red] ({360/\n * 9}:\radius) to [out= 170,in=-90,  loop] ({360/\n * 9}:\radius);
 \draw[red] ({360/\n * 10}:\radius) to [out= 170,in=-90,  loop] ({360/\n * 10}:\radius);
 \draw[red] ({360/\n * 11}:\radius) to [out= 170,in=-90,  loop] ({360/\n * 11}:\radius);
 \draw[red] ({360/\n * 12}:\radius) to [out= 170,in=-90,  loop] ({360/\n * 12}:\radius);
 \draw[red] ({360/\n * 13}:\radius) to [out= 170,in=-90,  loop] ({360/\n * 13}:\radius);
 \draw[red] ({360/\n * 14}:\radius) to [out= 170,in=-90,  loop] ({360/\n * 14}:\radius);
 \draw[red] ({360/\n * 15}:\radius) to [out= 170,in=-90,  loop] ({360/\n * 15}:\radius);

\draw[thick, blue] ({360/\n * 8}:\radius) to [out= 170,in=-90,  loop right ] ({360/\n * 8}:\radius);
\draw[thick, blue] ({360/\n * 0}:\radius) to [out= 170,in=-90,  loop right ] ({360/\n * 0}:\radius);

\node at ({360/\n * 12}:3cm)  {(\textit{b})};

 \end{tikzpicture}
  \caption{(Color online)  (a) Hanoi network  of  degree  three(HN3)   and  (b)  of degree  four(HN4)   with $16$ vertices.    One  self-loop(red)  has been added  at each vertex for lackadaisical quantum walks.}

\end{figure}
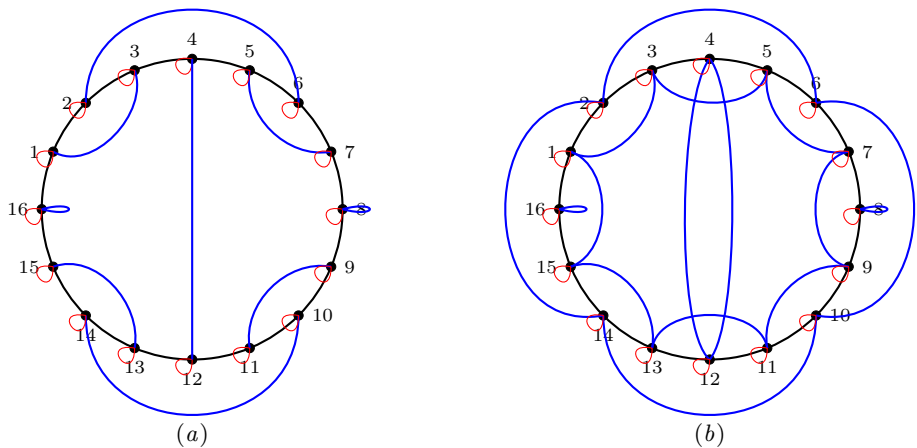

\section{ Hanoi network of degree three and four} \label{hanoi}
Hanoi network(HN) has been described  in detail  in  ref. \cite{boe}   as a network   generated from the    sequence of   numbered disks  of  the  Tower-of-Hanoi problem.  It can have   degree three and  degree four. 
A   one-dimensional   lattice   of size  $N=  2^n$ with  periodic  boundary conditions is the main structure of this network.     Each vertex   $1 \leq    x_v \leq  2^n$  has  two edges connected to its two nearest neighbors.   There are additional   long-range  edges, which give  the small-world  structure.  In figure 1  the   main structure  with $N=16$  has been shown in black color.  Long range edges  are shown in blue color.   We  have added  one   directed self-loop(red)  at each vertex of the Hanoi network for the purpose of  spatial  search  by  lackadaisical   quantum walks.

Each vertex   $x_v$   is located  at  position  $j$   in a specific hierarchy  denoted  by $i$, which obeys  the following  relation 
\begin{eqnarray}
x_v= 2^{i} \left(  2j +1  \right)\,,
\label{hier}
\end{eqnarray}  
where   the pair  $0 \le i \le n$  and   $0 \le j \le   j_{max}= \floor{2^{n-i-1} -  1/2}$ uniquely identifies the vertex  $x_v$.  There are  $N$ basis  states 
\begin{eqnarray}
|x_v \rangle  =   |i,  j \rangle\,,
\label{hier}
\end{eqnarray}  
expressed   in terms of   $x_v$ or   alternatively   in terms of  $i, j$.

Hanoi network of degree three(HN3)   has three edges  at each vertex.   Two of the edges which connect   $|x_v \rangle$   to   $|x_v -1  \rangle$  and  
$|x_v +1  \rangle$,  come from the  one-dimensional periodic lattice.  Third  edge    connects         vertices  in a     hierarchy, i.e., $|i, 0 \rangle$ is connected to  $|i, 1\rangle$,  $|i, 2\rangle$ is connected to  $|i, 3\rangle$, $\cdots$, and   $|i,  j_{max}-1 \rangle$ is connected to $|i,  j_{max} \rangle$.   For $i = n-1$ and $i =n$ the  third edge  forms  a directed  self-loop, i.e.,  $|n-1, 0\rangle$ and  $|n, 0\rangle$ connect to itself respectively.   From  fig. 1(a)   we can see that 
there are two black edges  and one blue long-range edge  at each vertex.   The  long-range  edge   at  $0$  and  $8$  forms  a directed   self-loop(blue).

Hanoi network of degree four(HN4)   has four edges, out of which two come from the one-dimensional periodic lattice as the HN3 case.  The rest of the two edges connecting  
$|i, j \rangle$   to   $|i, j -1  \rangle$  and  $|i, j +1\rangle$ respectively, for  $i \neq n-1, n$.  There is one  undirected    self-loop at  $|n-1, 0\rangle$ and  $|n, 0\rangle$  respectively.   From  fig. 1(b)   we can see that  there are two black edges  and two  blue long-range edges  at each vertex.   Two  long-range   edges   form an  undirected  blue   self-loop  at  $0$  and  $8$ respectively. 

\section{ Quantum walk   search on  Hanoi network} \label{qw}  
Quantum walks  is useful to   search for a target state  on Hanoi network  as   briefly mentioned  in the introduction.  For a  detail discussion on   efficient search in this network  using   modified Grover coin and   by  Tulsi's controlled quantum walk  see     refs.   \cite{portugal3,portugal4}. 
The  $N$-dimensional Hilbert space,   $\mathcal{H}_V$,     of     the  vertices of  Hanoi network      is      the  database in  quantum walks.   We impose periodic boundary conditions on each hierarchy  of the network.  We assume that one of the  basis states   of the Hilbert space  $\mathcal{H}_V$  is the target state.  Our task  is  to find out  the target state  with the help of quantum walk search. 
There are $d$-directions in which the quantum walker can move, which form the basis states for   the   coin space  $\mathcal{H}_C$.   For HN3 $d=3$ and  for HN4 $d=4$. 
The quantum walk,  which   is the rotation of the coin state followed by the shift operation on the vertex space,   takes  place on  the tensor product space  $\mathcal{H} = \mathcal{H}_C \times \mathcal{H}_V$.   An initial state of  the form 
\begin{eqnarray}
|\psi_{in}\rangle =    \frac{1}{\sqrt{d}} \sum^{d-1}_{x_c =0} |x_c\rangle  \otimes \frac{1}{\sqrt{N}} 
\sum^{N}_{x_v=1} |x_v \rangle \,,
\label{in}
\end{eqnarray}
is considered for the purpose of search by quantum walks, where   $|x_c \rangle$ and  $|x_v \rangle$ are  the basis states of the coin space and  vertex space respectively.  In analogy with the  classical random  walks,  in quantum case,  the evolution operator  as stated above  has  two components.  One for the toss of the coin, which can be simulated by unitary operator $C$ acted on the coin space and the other for the shift of the walker from one vertex to the next adjacent vertex, which can be done by  the shift operator $S$. Formally we can write the unitary  evolution operator as 
\begin{eqnarray}
 \mathcal{U} = S C\,, 
\label{uqw}
\end{eqnarray}
which acts repeatedly on the initial state to let it evolve.  To exploit this quantum walks to search for a target state we need to have a  way to distinguish  the target state   $|t_v \rangle$  from the rest of the basis  states.  Usually   it  is  done by using  two different coins   for   the target state and   for  the  rest of the    non-target states respectively.   So, let us   use  $C$  for the non-target states and  $- C$  for the target state, which can be compactly expressed as 
\begin{eqnarray}
\mathcal{C}=  C  \otimes \left( \mathbb{I} -   2 |t_v \rangle \langle t_v | \right)\,. 
\label{qc}
\end{eqnarray}
 Note that $\mathcal{C}$  acts on the tensor product space   $\mathcal{H}$  as opposed to  $C$   which only acts on the Hilbert space of the coin  $\mathcal{H}_C$.
 We choose $C$ to be the Grover coin prepared from the basis states   $|x_c \rangle$   of the coin space.

For the  quantum search on the  spacial region,  we use    flip-flop shift operator,   which   moves  the walker from one vertex to the other and inverts the direction of the walker at the same time.   For  HN3   it is of the form 
\begin{eqnarray} \nonumber
S_{HN3} &=&  \sum^{ N}_{x_v=1}  \left( |1\rangle \langle 0 | \otimes | x_v +1  \rangle \langle  x_v  | 
+  |0 \rangle \langle 1 | \otimes | x_v -1 \rangle \langle  x_v | \right)  \\ \nonumber
&+& \sum^{ n-2}_{i=0}   \sum^{j_{max}}_{j=0}  |2 \rangle \langle 2 | \otimes | i, j + (-1)^{j}  \rangle \langle  i, j | \\
&+&  |2 \rangle \langle 2 | \otimes  \left( | n-1 , 0  \rangle \langle  n-1, 0 |  +   | n , 0  \rangle \langle  n, 0 | \right)\,,
\label{fshift}
\end{eqnarray}
and  for  HN4    one more  edge is added  to each vertex, which  is incorporated  into the shift operator as 
\begin{eqnarray} \nonumber
S_{HN4} &=&  \sum^{ N}_{x_v=1}  \left( |1\rangle \langle 0 | \otimes | x_v +1  \rangle \langle  x_v  | 
+  |0 \rangle \langle 1 | \otimes | x_v -1 \rangle \langle  x_v | \right)  \\ \nonumber
&+& \sum^{ n-2}_{i=0}   \sum^{j_{max}}_{j=0}  \left(  |3 \rangle \langle 2 | \otimes | i, j + 1  \rangle \langle  i, j |  
+ |2 \rangle \langle 3 | \otimes | i, j - 1  \rangle \langle  i, j | \right) \\
&+& \left( |2 \rangle \langle 2 | +   |3 \rangle \langle 3 |  \right)\otimes  \left( | n-1 , 0  \rangle \langle  n-1, 0 |  +   | n , 0  \rangle \langle  n, 0 | \right)\,.
\label{fshift1}
\end{eqnarray}
Note that   states $|0 \rangle,  |1 \rangle$  are the regular edges  and    $|2 \rangle,  |3 \rangle $   are the long range edges  of the networks, which form the coin space.  HN3  has only one  long range edge   $|2 \rangle$    and   HN4 has  two long range edges   $|2 \rangle,  |3 \rangle $.

The initial state   $|\psi_{in }\rangle $  is  evolved    by    repeated application  of    the   evolution operator   $\mathcal{U} =  S\mathcal{C}$  until  the  probability of the  target state is  reached  to its maximum. 
The final state   thus   obtained 
\begin{eqnarray}
|\psi_{f }\rangle =    \mathcal{U} ^{T} |\psi_{in }\rangle  \,,
\label{fi}
\end{eqnarray}
is  not aligned with  the target state with high   fidelity  for the quantum  search on Hanoi network.      In fact,       $ |\langle t_v |\psi_{f }\rangle|^2    << 1 $,   which implies that we need amplitude  amplification   technique of Grover   to amplify the   probability amplitude of  the  target state  to  $\mathcal{O}(1)$  in  the evolving  state.

\begin{figure}[h!]
  \centering
     \includegraphics[width=1.0\textwidth]{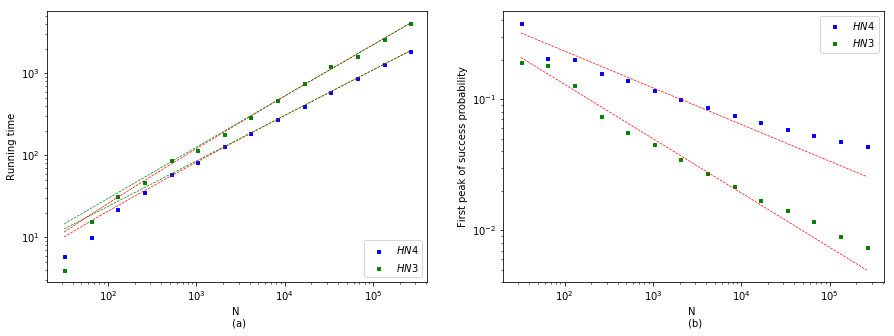}
          
       \caption{(Color online) (a) Running time and  (b) corresponding  success probability  to find  the target state  $|4 \rangle$   as a function of the number of vertices $N= 2^n, n = 5, 6, \cdots, 18$   of   HN3(green squares)  and  HN4(blue squares)  network .  Red dash curves are the best fits.  Green dashes  in figure (a) are  another best fits.}

\end{figure}

In  figure 2   we present   the result of the  numerical  simulation in log-log-scale  for the  quantum search  of  the target  state   $|4 \rangle$   on  Hanoi network     by regular quantum walks(without self-loop). 
Running  time    to obtain the first peak of the success probability   as a function of the   number of vertices  $N$    is shown in fig. 2(a)  and  its corresponding  success probability of the first  peak   is plotted  in  fig. 2(b).     Green   and  blue  squares  are the numerical data for   HN3 and   HN4  network  respectively.    From the  curve(red dash)  fitting,   we see that for  HN3 the  running time and  the success probability   scale   respectively  as    $\mathcal{O}(N^{0.57913867}\sqrt{\log N})$  and   $\mathcal{O}(N^{-0.41444274})$.    And for  HN4 the running time and  the success probability  scale  as   $\mathcal{O}(N^{0.50823082}\sqrt{\log N})$  and   $\mathcal{O}(N^{-0.28004801})$.    After  an  appropriate number of  amplitude amplifications,   the   running time for   HN3 and HN4  are respectively  given by 
\begin{eqnarray} \nonumber
T_{HN3} &\approx&   \mathcal{O}(N^{0.79} \sqrt{\log N}) \,,\\  
T_{HN4}  &\approx&   \mathcal{O}(N^{0.65}\sqrt{\log N}) \,.
\label{timeamp}
\end{eqnarray}
This result  indicates that the long-range edges  facilitate    the   search by the quantum walks on Hanoi network.   In the above analysis we  assumed    $aN^b \sqrt{\log N}$  and  
$c/N^d$  as the form of the functions for the running time and  the success probability   respectively  for the curve fitting, where  the constants  $a, b, c, d$ are  evaluated by the curve fit.  
If  we  consider $aN^b$  as   the form for the running  time, then  we obtain   green  dash curves in figure 2(a)  as the curve fit,    which  although  does not fit well for small network sizes but for large network sizes   fit  well.   In this case     $T_{HN3}  \approx  \mathcal{O}(N^{0.83}) $  and  $T_{HN4}  \approx \mathcal{O}(N^{0.69}) $  are the running times  with   $\mathcal{O}(1)$ success probability.

\section{LQW  search  on Hanoi network} \label{lack}
In this section,  we discuss  how lackadaisical quantum walks  can be  exploited to search  a target state on Hanoi network. It has been observed  that  the lackadaisical quantum walks  can   search target states in two-dimensional lattice  in    $\mathcal{O} (\sqrt{\frac{N}{M} \log \frac{N}{M}})$  time steps \cite{giri1}.   In one-dimensional  periodic lattice also it can be efficiently used   for the search algorithm. 

In this approach,   a self-loop of weight  $l$  is attached to each vertex of the   Hanoi network   as shown  by the red  self-loops in figure 1.  The initial state  becomes 
\begin{eqnarray}
|\psi_{in}\rangle =   |\psi_c\rangle   \otimes \frac{1}{\sqrt{N}} 
\sum^{N}_{x_v=1} |x_v \rangle \,,
\label{inl}
\end{eqnarray}
where  the initial state of  the coin  is of the form 
\begin{eqnarray}
|\psi_c\rangle  =  \frac{1}{\sqrt{d + l}}   \left(\sum^{d-1}_{x_c =0}   |x_c\rangle  + \sqrt{l}| \mbox{loop} \rangle\right)\,.
\label{coinl}
\end{eqnarray}
\begin{figure}[h!]
  \centering
     \includegraphics[width=1.0\textwidth]{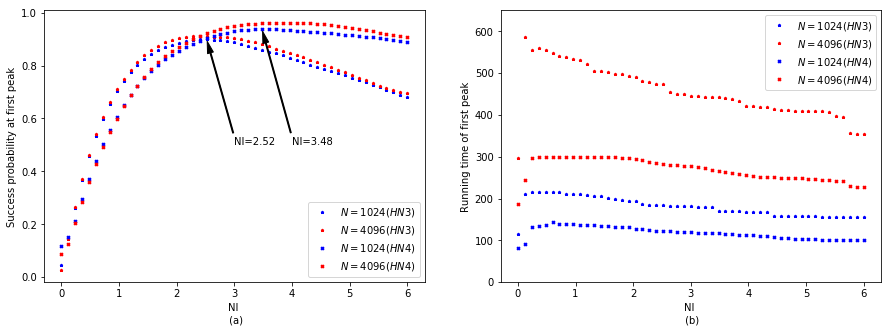}
          
       \caption{(Color online) (a) Success probability of the  target state  $|4 \rangle$   and  (b) corresponding   running time   as a function of   $Nl$       
       for  the  Hanoi network of degree three(star curves)   and  four(cross curves)  for     $N=1024, 4096$. }

\end{figure}
Note that  the state  $|\mbox{loop} \rangle$  in the coin space is responsible for the  laziness of  the quantum walk.   In order to search by lackadaisical  quantum walks     we first need   to   rotate the    coin state   $|\psi_c\rangle$     by   a coin operator, which is   followed by    flip-flop shift operator  acting on   the vertex state.   The Grover diffusion operator   for  the rotation of  the coin state is given by 
\begin{eqnarray}
C= 2 |\psi_c\rangle \langle \psi_c| - \mathbb{I}_{d+1}\,, 
\label{hcgrov}
\end{eqnarray}
Note that  for the  spatial search  the modified coin   (\ref{qc})  can be obtained   replacing  $C$    by      eq.  (\ref{hcgrov}).   The shift operators  are also  modified because of the  self-loop term.  For the Hanoi network  of degree three and four,   it can be    respectively    written  as 
\begin{eqnarray} \nonumber
\widetilde{S_{HN3}}   =  S_{HN3}  +  |\mbox{loop} \rangle \langle \mbox{loop} | \otimes  \sum_{x_v=1}^{N}  | x_v  \rangle \langle  x_v | \,, \\
\widetilde{S_{HN4}}   =  S_{HN4}  +  | \mbox{loop} \rangle \langle \mbox{loop} | \otimes  \sum_{x_v=1}^{N}  | x_v  \rangle \langle  x_v | \,.
\label{fshiftm}
\end{eqnarray}
Note that the self-loop terms in the shift operator retain  the probability amplitude at the same vertex. It works  in favor of increasing the efficiency of   the search algorithm  when the shift operator acts on the target state. However, it works against the increase of efficiency  of   the search algorithm when the shift operator acts on the non-target states.  Therefore we need to balance it  in such a way that we obtain an overall increase of the efficiency   of the search  algorithm, which can be done by suitably controlling the weight $l$ of the self-loop.  

The initial states are   evolved  by the  repeated  application of the  unitary operators  
\begin{eqnarray} \nonumber
\mathcal{U}_{HN3}  &=&  \mathcal{C}\widetilde{S_{HN3}}   \,, \\
\mathcal{U}_{HN4}  &=& \mathcal{C}\widetilde{S_{HN4}}   \,.
\label{unitm}
\end{eqnarray}
and we obtain the final states  as
\begin{eqnarray} \nonumber
|\psi_{HN3}\rangle  &=&    \mathcal{U}_{HN3} ^{T} |\psi_{in }\rangle  \,,\\  
|\psi_{HN4}\rangle  &=&    \mathcal{U}_{HN4} ^{T} |\psi_{in }\rangle \,.
\label{fisel}
\end{eqnarray}

\begin{figure}[h!]
  \centering
     \includegraphics[width=1.0\textwidth]{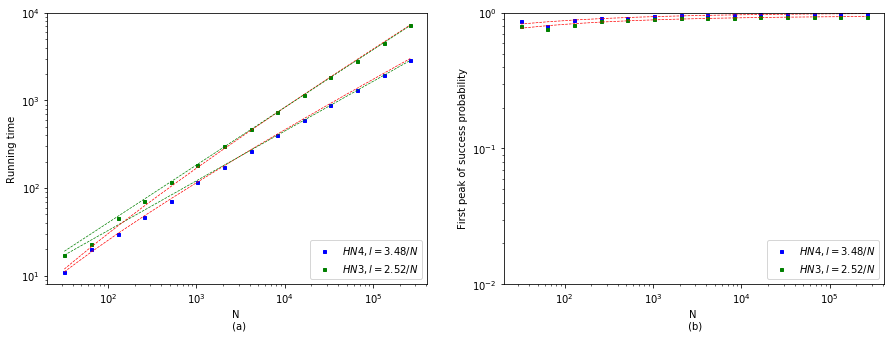}
          
       \caption{(Color online)(a) Running time  and   (b) success probability of the  target state  $|4 \rangle$ as a function of the lattice size $N= 2^n, n= 5, 6, \cdots, 18$  for  the  Hanoi network of degree three(HN3)  and degree four(HN4) respectively.  Red dashes(a) are the   best fit with the numerical data.  Green dashes in figure (a) are  another best fits.}

\end{figure}

In order to  exploit  the effect of  the  self-loop in  search algorithm,  we need to  obtain the optimum  values of the self-loop parameter $l$ for which the first peak of the success probability is maximum.    We plot  the success probability  as  a function of   $Nl$   in figure 3(a)   and  the corresponding  running  time  in  figure 3(b)  for two different   lattice sizes   $N=1024$(blue  curves)   and  $N= 4096$(red curves).    Star curves  refer  to the data for  HN3  network and   thick cross  curves refer to the data for HN4 network.   We   observe that the  curves for the success probability  maximize at  $l= 2.52/N$  for HN3 and   at   $l= 3.48/N$  for HN4,  which will be  held fixed  from now onward   for the  analysis of  searching  for a   target  state.

In figure 4  the numerical data for   the search of  the target state  $|4\rangle$   on  Hanoi network of degree three(HN3)  and degree four(HN4)   by lackadaisical quantum walks   has been  presented in log-log-scale.    Green and blue  square  curves  represent   HN3 and  HN4  respectively.   The success probabilities(fig. 3(b))  approach to a constant  $\mathcal{O}(0.94820898)$  and $\mathcal{O}(0.99439784)$ for  HN3 and HN4 respectively  for large $N$.   From the best fit(red dashes) of the running time curves(fig. 2(a))  we  obtain  
\begin{eqnarray} \nonumber
T_{HN3} &\approx&   \mathcal{O}  (N^{0.57}\log N)  \,,\\  
T_{HN4}  &\approx&   \mathcal{O} (N^{0.50}\log N)  \,, 
\label{timefit}
\end{eqnarray}
as the  running  time  needed to  perform the quantum spatial search  on HN3 and HN4 respectively.   The  results are valid for any choices of the target vertex  except  for   $x_v= 2^{n-1}$ and   $2^n$, which can not be found  by (lackadaisical)  quantum walk search.    These two vertices are special, because  the long-range  edges  form a self-loop, which prohibits   the flow of  probability flux  from other neighboring   vertices to  these two vertices   through the  long-range edges.  

Note that in the above analysis we  assumed    $aN^b \log N$  as the form of the function  for the running time  for  curve fitting, where  the constants  $a, b$ are  evaluated by the curve fit as done in the previous section.  
If  we  instead  consider $aN^b$  as   the form for the running  time, then  we obtain   green  dash curves in figure 4(a)  as the curve fit,    which    does not fit well for small network sizes but for large network sizes   fit  well.   In this case     $T_{HN3} \approx  \mathcal{O}(N^{0.66}) $  and  $T_{HN4} \approx \mathcal{O}(N^{0.57}) $  are the running times  with   $\mathcal{O}(1)$ success probability.

\section{Conclusions} \label{con}
Quantum algorithm    for  search  on  a  physical database is usually  based  on   quantum walks.  In  $d \geq 3$-dimensional periodic lattice  the time complexity of  the quantum walk  search algorithm   can reach to  the  optimum value  of $\mathcal{O}(\sqrt{N})$. However, for $d \leq 2$  the situation is different. For example,  on   $2$-dimensional periodic Cartesian lattice with degree four one can find a target state in  $\mathcal{O}(\sqrt{N \log N})$ time steps.   In one-dimensional periodic lattice,  regular quantum walk search is not efficient. 
However using  lackadaisical quantum walk followed by the   amplitude amplification  one can  obtain  a target state  in    $\mathcal{O}\left(N\right)$ time steps   with    $\mathcal{O}(1)$ success   probability.  Lackadaisical quantum walks is also used in two-dimensional periodic lattice to search for $M$ targets, which needs   $\mathcal{O}\left(\sqrt{\frac{N}{M}\log \frac{N}{M}}\right)$ time steps.  This numerical result has been tested   upto $M=6$  in ref. \cite{giri1}.

Given the success of the lackadaisical quantum walks in searching target states in one- and two-dimensional periodic lattice we in this article considered another graph known as Hanoi network.   We also  studied  the quantum search  on this network  using  regular quantum walk followed by  amplitude amplification. 
Hanoi network  is   different from the one-dimensional periodic lattice, because it has extra long-range edges on top of the regular two-edges at each vertex.  It has been observed   that these  extra long-range edges enhance    the efficiency of the quantum   walk  search. 
Our  numerical simulation suggests that the running  time for the regular quantum walks  followed by amplitude amplification    for  the  Hanoi network of degree three and four scales  as  $\mathcal{O}\left(N^{0.79} \sqrt{\log N}\right)$   and     $\mathcal{O}\left(N^{0.65} \sqrt{\log N}\right)$, which are in the class of functions    $\mathcal{O}(N^{0.83}) $  and  $\mathcal{O}(N^{0.69})$   respectively.   
Using   the lackadaisical quantum walks  as the search  algorithm  is even more efficient.   It has been shown   by   the curve fitting  in  figure 4(a)  that  the running time scales  as   $\mathcal{O}\left(N^{0.57}\log N\right)$   and   $\mathcal{O}\left(N^{0.50}\log N\right)$  for HN3 and HN4, which  are in the class of functions 
 $\mathcal{O}(N^{0.66}) $  and  $\mathcal{O}(N^{0.57}) $   respectively.   The class of function  $\mathcal{O}(N^{0.66})$   is in  agreement  with the result   $N^{d_s}$ obtained  in ref.  \cite{boe1}  for  the running time  of  HN3  network, where $d_s= 0.652879$ is the spectral dimension of the network.

\section*{Acknowledgements} 
P. R. Giri is supported by  International Institute of Physics, UFRN, Natal, Brazil.  V.   Korepin is  grateful to SUNY Center of Quantum Information Science at Long Island for support.  We thank     Chen-Fu Chiang for his comments/suggestions.


\end{document}